\documentclass[12pt,a4paper]{article}

\usepackage{amsmath, amssymb, amsthm}
\usepackage{graphicx}
\usepackage{tikz}
\usepackage{setspace}
\usepackage{geometry}
\usepackage{titlesec}
\usepackage{hyperref}
\usepackage{footnote}
\usepackage{endnotes}
\usepackage{natbib}
\usepackage{lmodern} 
\usepackage[T1]{fontenc}
\usepackage{caption}
\usepackage{float}
\usepackage{bm}

\usepackage[utf8]{inputenc}
\usepackage{graphicx}
\usepackage{amsmath}
\usepackage{setspace}
\usepackage{geometry}
\usepackage{times}
\usepackage{hyperref}
\usepackage{natbib}
\usepackage{amsfonts}
\usepackage{authblk} 
\titleformat{\chapter}[hang]{\LARGE\bfseries}{\thechapter\quad}{0pt}{}

\begin{document}














\title{\textbf{A Review of Statistical and Machine Learning Approaches for Coral Bleaching Assessment}}
\author{Soham Sarkar and Arnab Hazra}
\affil{Department of Mathematics and Statistics, Indian Institute of Technology Kanpur, Kanpur 208016, India}
\date{}



\maketitle
\doublespacing

\begin{abstract}
Coral bleaching is a major concern for marine ecosystems; more than half of the world's coral reefs have either bleached or died over the past three decades. Increasing sea surface temperatures, along with various spatiotemporal environmental factors, are considered the primary reasons behind coral bleaching. The statistical and machine learning communities have focused on multiple aspects of the environment in detail. However, the literature on various stochastic modeling approaches for assessing coral bleaching is extremely scarce. Data-driven strategies are crucial for effective reef management, and this review article provides an overview of existing statistical and machine learning methods for assessing coral bleaching. Statistical frameworks, including simple regression models, generalized linear models, generalized additive models, Bayesian regression models, spatiotemporal models, and resilience indicators, such as Fisher’s Information and Variance Index, are commonly used to explore how different environmental stressors influence coral bleaching. On the other hand, machine learning methods, including random forests, decision trees, support vector machines, and spatial operators, are more popular for detecting nonlinear relationships, analyzing high-dimensional data, and allowing integration of heterogeneous data from diverse sources. In addition to summarizing these models, we also discuss potential data-driven future research directions, with a focus on constructing statistical and machine learning models in specific contexts related to coral bleaching. 

\textbf{Keywords}: Climate change, Coral bleaching, Machine learning, Predictive ecology, Statistical modeling, Coral resilience.

\end{abstract}

\section*{List of Abbreviations}

\begin{itemize}
    \item[$\blacksquare$] ANN: Artificial neural network
    \item[$\blacksquare$] BRT: Boosted regression trees 
    \item[$\blacksquare$] CPT: Conditional probability table
    \item[$\blacksquare$] DHW: Degree heating week
    \item[$\blacksquare$] FI: Fisher's information
    \item[$\blacksquare$] GAM: Generalized additive model
    \item[$\blacksquare$] GBR: Great Barrier Reef
    \item[$\blacksquare$] GLM: Generalized linear model
    \item[$\blacksquare$] GMRF: Gaussian Markov random field
    \item[$\blacksquare$] GR: Generic richness
    \item[$\blacksquare$] ML: Machine learning
    \item[$\blacksquare$] NB: Negative binomial
    \item[$\blacksquare$] NOAA: National Oceanic and Atmospheric Administration 
    \item[$\blacksquare$] RF: Random forest
    \item[$\blacksquare$] SST: Sea surface temperature
    \item[$\blacksquare$] SVM: Support vector machine
    \item[$\blacksquare$] WQ: Water quality
    \item[$\blacksquare$] ZINB: Zero-inflated negative binomial
\end{itemize}

\section{Introduction}

In coral bleaching, symbiotic dinoflagellates and their associated photosynthetic pigments are expelled from coral tissues, causing the corals to turn white. A variety of environmental stressors, including fluctuations in water temperature, light intensity, and salinity, are the primary factors for coral bleaching. Bleached corals experience increased physiological stress, making them more susceptible to malnutrition, disease, and mortality \citep{vanOppenLough2018}. Increased SST due to anthropogenic climate change is the primary driver of coral bleaching  \citep{donner2005global, hughes2017global, sully2019global, johnson2023global}. Marine heatwaves produce more reactive oxygen species within the symbiotic algae, which leads to cellular oxidative damage, and as a result, host tissues expel the algae. Mass bleaching events have significantly increased in frequency, intensity, and geographical scale over the past three decades \citep{hughes2017global, sully2019global}. Bleaching intensities show a high heterogeneity across reefs and species under similar thermal stress \citep{heron2016validation, johnson2023global} due to the fluctuations of local environmental conditions, coral compositions, and adaptive changes that include genetic adaptations, and shifts in symbiont assemblages \citep{lachs2023emergent}. According to evidence from the Pacific and Indian Oceans, coral assemblages may become more thermally tolerant through community turnover or physiological adaptation-- approximately $0.1^\circ$C per decade \citep{lachs2023emergent}.

Recent literature has identified some potential ecological factors that may reduce the chances of coral bleaching. For instance, proximity to mangroves may help reduce the prevalence of bleaching under thermal stress, likely due to physical shedding, altered water chemistry, and co-tolerant coral-symbiont assemblages \citep{johnson2023global}. Moreover, by controlling the interactive effects of temperature and light, physical variables like turbidity and local hydrodynamics mitigate the severity of bleaching in specific microhabitats \citep{sully2019global}. Due to a high spatial and temporal heterogeneity, probabilistic modeling of coral bleaching requires appropriate quantitative frameworks that integrate environmental covariates, spatial structure, and temporal dynamics. Traditional SST-based deterministic models have limited capacity to explain the nonlinear and high-dimensional nature of coral responses \citep{heron2016validation}. As a result, probabilistic and data-driven frameworks, including Bayesian hierarchical models, spatiotemporal models, and machine learning algorithms, provide the necessary tools to improve prediction accuracy and uncertainty quantification \citep{bainbridge2019use, peterson2020monitoring, hazra2021estimating}. For example, Bayesian hierarchical models have been successfully used to produce real-time bleaching risk indices and forecasts during the 2015--2016 GBR bleaching event, which assimilate high-frequency in-situ data with environmental stressors such as temperature, light, and wind \citep{bainbridge2019use}. Similarly, allowing the fusion of citizen science and professional monitoring data, proposed spatiotemporal models improve the spatial resolution and uncertainty quantification of coral cover predictions by more than $40\%$ \citep{peterson2020monitoring}. By incorporating local environmental conditions, spatial influences, and the inherent resilience of coral, the data fusion methods provide improved inference and prediction. Further, they allow the identification of latent patterns within complex datasets and the development of predictive models with direct applicability to conservation and management efforts. Remote-sensing products, such as the NOAA Coral Reef Watch high-resolution ($5$ km) DHW and hotspot datasets, have further improved large-scale bleaching surveillance \citep{heron2016validation}. However, these metrics cannot capture local-scale ecological and environmental heterogeneity, which necessitates further integration of statistical and ML models with such data. This integration enables researchers to translate global thermal stress metrics into site-specific risk assessments, thereby enhancing the operational relevance of these assessments for reef management and conservation planning.

In this article, we review statistical and machine learning methodologies used to address various challenges related to coral bleaching. We categorize the most frequently used models in the existing literature, with a particular focus on how local environmental factors influence the dynamics of coral bleaching. By summarizing these approaches, we aim to present a comprehensive understanding of the mathematical frameworks in coral bleaching research and to highlight their scientific and practical relevance. Furthermore, this review presents key insights from these models, identifies critical research gaps, and outlines future directions for developing models using new statistical and machine learning methodologies that can capture the complex, multi-scalar nature of coral reef responses to environmental stressors.

The paper is organized as follows. In Section \ref{sec:stat_methods}, we discuss different statistical approaches for coral bleaching assessment, including frequentist regression models in Section \ref{subsec:regression}, Bayesian regression and state-space models in Section \ref{subsec:bayesian}, Bayesian spatiotemporal models in Section \ref{subsec:bayesian_spatiotemporal}, Bayesian causal inference in Section \ref{subsec:bayesian_causal}, and some advanced statistical metrics in Section \ref{subsec:statistical_metrics}. Section \ref{sec:machine_learning} discusses machine learning approaches, including supervised learning techniques in Section \ref{subsec:supervised} and unsupervised learning techniques in Section \ref{subsec:unsupervised}. In Section \ref{sec:future}, illustrating with an example, we briefly discuss the necessity of building data-specific models instead of analyzing the data using readily available tools. Section \ref{sec:conclusion} concludes.

\section{Statistical methods}
\label{sec:stat_methods}
The literature on coral bleaching covers a range of methodological frameworks for predicting and understanding bleaching phenomena. Recent studies have employed several analytical paradigms, including classical statistical approaches and Bayesian hierarchical frameworks, to capture the complex, nonlinear, and multi-scalar nature of bleaching events \citep{donner2005global, heron2016validation, sully2019global,  bainbridge2019use, johnson2023global, peterson2020monitoring}. Traditional statistical models have been heavily used in establishing empirical relationships between environmental covariates such as light, temperature, and depth, and observed bleaching responses \citep{heron2016validation, sully2019global}. On the other hand, to account for multi-source and spatially structured data, integration of Bayesian and spatiotemporal models has gained increasing popularity, offering better uncertainty quantification \citep{bainbridge2019use, johnson2023global}. Accordingly, we can broadly classify the models utilized to study coral bleaching into the following categories.

\subsection{Regression-based models for inference and prediction}
\label{subsec:regression}
This section covers models that are under the classical statistical paradigm and data-driven algorithmic methods. Frequentist models typically rely on single-point estimates and make inferences based on distributional assumptions of the errors \citep{donner2005global, heron2016validation, sully2019global}. These frameworks include regression-based and generalized linear models that quantify the relationships between the bleaching intensities and their environmental stressors, such as temperature, light, depth, and habitat variability \citep{heron2016validation, sully2019global}.

\subsubsection{Linear and nonlinear regression}

When predicting future events, we primarily consider a linear relationship between the response variable and the covariates, incorporating them into a linear model that adheres to standard assumptions. \cite{donner2005global} discuss a similar model to assess coral bleaching globally via a statistical downscaling approach. Instead of linking bleaching intensity directly, the authors model the relationship between the local SST and the Global Circulation Model (GCM)-resolution SST by a linear regression model of the form
\begin{equation}
Y = \beta_0 + \beta_1 X + \varepsilon,
\end{equation}
where $Y$ denotes monthly local SST, $X$ denotes spatially averaged monthly SST for the entire GCM grid cell containing that reef cell, $\beta_0$ and  $\beta_1$ are the regression coefficients calculated for each specific cell using historical satellite data, and $\varepsilon$ denotes the additive error term.

In certain situations, a linear systematic component may fail to adequately capture the relationship between the response variable and its predictors. In such cases, it becomes necessary to consider nonlinear modeling frameworks to represent these complex associations more accurately. \cite{heron2016validation} explore this idea by incorporating nonlinear relationships into their modeling framework, thereby enhancing the flexibility and predictive capability of the model for estimating the percentage of bleached coral. Their study emphasizes the importance of validating the performance of NOAA’s newly developed high-resolution (5 km) satellite products designed to monitor coral reef thermal stress. Although these tools represent significant technological advancements, they lack direct evaluation of how effectively their thermal stress metrics, like DHWs, correspond with observed bleaching events. To address this limitation, the authors begin their analysis by employing simple linear regression and pairwise correlation analyses to assess the relationship between satellite-derived DHW and observed coral bleaching percentages. Through multiple correlation analyses, they identify the most influential predictors from a broad set of variables, including DHW and GR. Based on these findings, two predictive models for the percentage of coral bleaching are: an interactive linear model and a nonlinear interactive model. The general formulations of these models are as follows.
\paragraph{Interactive linear model:} Here we model the coral bleaching intensity $Y$ using the DHW $X_1$ and GR $X_2$ by
\begin{equation}\label{eq:interactive_linear}
Y = \beta_0 + \beta_1 X_1 X_2 + \varepsilon,
\end{equation}
where $\beta_0$ and $\beta_1$ are the regression coefficients and $\varepsilon$ is the random error term.

\paragraph{Nonlinear interactive model:} Here we alternatively model the coral bleaching intensity $Y$ using the DHW $X_1$ and GR $X_2$ by
\begin{equation}\label{eq:interactive_nonlinear}
Y = \beta_0 + \beta_1 X_1^{d} X_2^{g} + \varepsilon,
\end{equation}
where $\beta_0$ and $\beta_1$ are the regression coefficients, $\varepsilon$ is the random error term, and $d$ and $g$ determine the order of nonlinearity.

\subsubsection{Generalized linear models}
In many instances, both linear and nonlinear regression models may prove inadequate or inappropriate for describing complex ecological relationships, thereby necessitating the use of a more general modeling framework. In such contexts, GLMs offer a versatile and robust approach. \cite{safaie2018high} employ a specific type of GLM—ordinal logistic regression—to examine how interactions among various explanatory variables influence the relative log-odds of a given bleaching prevalence score. In ecological studies, datasets are frequently represented using semi-quantitative variables, where relative differences among values carry significant meaning. For such data structures, ordinal logistic regression provides a suitable analytical framework. In the context of coral bleaching, the prevalence scores are standardized into an ordinal variable consisting of four ranked categories reflecting increasing severity of bleaching. This modeling approach has been applied widely to assess how environmental and physiological factors influence reef ecosystem responses following disturbances. The parameters of the model are estimated using maximum likelihood estimation, which quantifies the relative log-odds of observing a bleaching prevalence score less than or equal to a given level compared to the odds of observing a higher category. Recognizing that multiple thermal stress and environmental components can drive coral bleaching responses, \cite{safaie2018high} consider 20 experimental variables grouped into eight categories: (1) depth, (2) background oceanographic conditions, (3) cumulative thermal stress, (4) acute thermal stress, (5) thermal trajectory, (6) heating rate, (7) high frequency temperature variability, and (8) shape of the distribution of high frequency temperature variability. Using these eight explanatory variable classes and the bleaching prevalence score (ranging from 1 to 4) as the response variable, the study formulates an ordinal logistic regression model, which can be expressed as follows:
\begin{equation}
\log \left( \frac{\mathrm{P}(Y_i \leq j)}{\mathrm{P}(Y_i > j)} \right) = c_j + \beta_1 z_{i1} + \beta_2 z_{i2} + \cdots + \beta_p z_{ip},
\end{equation}
where $\mathrm{P}(Y_i \leq j)$ is the probability of observing bleaching in category $j$ or a less severe category, $\mathrm{P}(Y_i > j)$ is the probability of observing bleaching in a more severe category, $c_j$ is the model intercept for each category $j$, $\beta_k$'s are model coefficients that represent the effect of the explanatory variables, and $z_{ik}$'s are standardized independent variables (e.g., diurnal temperature range, depth, etc.) for the $i$-th observation.

The version of GLM may vary depending on the problem statement we are interested in. Suppose, in a broader context, we may want to predict binary outcomes, such as whether a bleaching event occurs or not, even on a small scale. \cite{kumagai2018high} mainly compare the GLM method and the Random Forest method. The purpose of using GLM here is to predict a binary outcome of bleaching events. We can conceptualize the model through the following example.

\paragraph{A two-stage model example:}
The mathematical formulation links the probability of bleaching to a linear combination of predictor variables. For a single predictor like SST, the formula for the bleaching probability $P$ is
\begin{equation}
P = \text{logistic}(\beta_0 + \beta_1 \, \text{SST}),
\end{equation}
where $\text{logistic}(x) = 1 / (1 + \exp[-x])$, and $\beta_0$ and $\beta_1$ are regression coefficients.

\cite{berkelmans2004comparison} use logistic regression for prediction, whose primary objective is to determine which specific satellite-derived SST variable is the most effective predictor of large-scale coral bleaching patterns on the Great Barrier Reef. They experiment in two stages. GAMs have been used as an exploratory tool to compare the predictive power of a wide range of different SST variables, where they are used to capture complex nonlinear relationships between a predictor variable (SST) and a response variable (bleaching) without assuming any predefined relationship between these two. Using this tool, the authors claim that maximum SST over a 3-day period is the best predictor variable in this setup. The authors then opt for logistic regression as a predictive model to quantify the sensitivity of the reef to temperature changes. In the second stage, the paper uses a logistic regression model to predict $P$ based on a given SST-based covariate value $x$. The core equation for the probability of bleaching is given as
\begin{equation}
P = \frac{1}{1 + \exp[-(a+bx)]},
\end{equation}
where $P$ is the probability that a reef will bleach, $x$ is the value of the predictor variable (in the final model, this was the maximum 3-day temperature, $\text{max3d}$), $a$ is the intercept of the model, and $b$ is the slope or coefficient for the predictor variable, which determines the steepness of the curve. The authors also express this relationship in terms of the \textit{odds} of bleaching by
\begin{equation}
\text{Odds} = \frac{P}{1 - P} = \exp[a+bx].
\end{equation}
A key part of the mathematical analysis is the calculation of the \textit{odds ratio}, which describes how the odds of bleaching change for a one-unit increase in the predictor variable. It is calculated from the slope coefficient $b$ as $\text{Odds Ratio} = \exp[b]$. In the paper, the estimated slope is $b = 1.76$ and therefore, the estimated odds ratio is calculated as $\exp[1.76] \approx 5.8$; this is the mathematical basis for their conclusion that a $1^{\circ}$C increase in temperature increases the odds of a reef bleaching by a factor of approximately 5.8.

\subsubsection{Analysis of covariance (ANCOVA)}

In ecology, it can be challenging to detect and quantify the impact of major disturbances due to the scarcity of data with high resolution and ecologically meaningful spatial extent. \cite{kopecky2023quantifying} discuss such problems and a potential solution. Although they develop a novel technological workflow, the final analysis step employs an ANCOVA model to compare the 2D and 3D metrics, with the primary focus being to test whether the 3D metric captures a greater rate of coral loss than the 2D metric. The model can be represented as
\begin{equation}
Y = \beta_0 + \beta_1 A + \beta_2 T + \beta_3 (A \times T) + \varepsilon,
\end{equation}
where $Y$ denotes the Live Coral Area, the response variable (measured surface area of live coral), $A$ denotes Area Type, a categorical predictor with two levels, namely \textit{2D Planar Area} and \textit{3D Surface Area}, $T$ denotes Time Point, a categorical predictor with two levels- \textit{Pre-bleaching 2018} and \textit{Post-bleaching 2019}, $(A \times T)$ denotes the interaction term between Area Type and Time Point, and $\varepsilon$ denotes the random error term.

\subsection{Bayesian approaches}
\label{subsec:bayesian}
Coral reef datasets are inherently complex, characterized by spatial and temporal dependencies, hierarchical structures, and non-standard statistical properties. As the analytical challenges in this domain continue to grow, researchers increasingly seek flexible modeling frameworks that extend beyond classical approaches. This shift has motivated the adoption of Bayesian methods, which provide a coherent framework for addressing model uncertainty and incorporating complex structures through the posterior distribution. In the existing literature, various Bayesian approaches have been proposed to address problems related to coral bleaching. Broadly, these modeling strategies can be categorized into the following classes.

\subsubsection{Bayesian generalized linear mixed models}
The literature has consistently demonstrated that thermal stress is a major driver of coral bleaching. However, the bleaching patterns often exhibit substantial spatiotemporal variability, influenced by geographic differences in both temperature anomalies and coral reef resilience. These spatial and temporal heterogeneities can lead to discrepancies between predicted bleaching patterns and actual field observations. \cite{sully2019global} address this challenge by employing generalized linear mixed models within a Bayesian framework, providing a more flexible structure to account for complex dependencies in the data. Their dataset comprises observations collected from thousands of coral reef sites, many of which are spatially clustered within distinct eco-regions. Because observations from nearby locations are often correlated and several sites are surveyed repeatedly, the authors incorporate random effects to account for within-site dependence. Furthermore, the response variable exhibits a zero-inflated distribution, prompting the adoption of a negative binomial model with a log-link function to accommodate overdispersion and excess zeros in the bleaching data. In a related study, \cite{johnson2023global} employ a similar modeling approach to investigate whether the proximity of mangroves exerts a protective influence against coral bleaching. To evaluate this hypothesis, they formulate a Bayesian generalized linear mixed model using a negative binomial distribution with a log-link function as the base model. The general structure of the model is expressed as follows.

For each observation $Y_i$, coral bleaching is modeled as arising from a series of Bernoulli processes with success probabilities $p_i$, which collectively follow an NB distribution to accommodate overdispersion and excess zeros in the data. A log-link function is employed to relate the mean bleaching response to the set of explanatory variables. Accordingly, the model can be formulated as $Y_i \sim \text{NB}(p_i, r)$, where
\begin{equation}\label{eq:NB}
\log(p_i) = \beta_0 + \beta_1 x_{i1}^* + \beta_2 x_{i2}^* + \cdots + \beta_n x_{in}^* + A_{\bm{s}(i)},
\end{equation}
with $\mathrm{E}(Y_i) = p_i$ and $\mathrm{Var}(Y_i) = p_i + p_i^2/r$, $\beta_0$ denotes the intercept, $\beta_j$ denotes the coefficient for the $j^{th}$ environmental covariate, $j = 1, 2, \ldots, n$, and $x_{ij}^*$ denotes the standardized covariate for the $i^{th}$ observation. The standardized covariates are defined as
\begin{equation}
\nonumber x_{ij}^* = \frac{x_{ij} - \text{mean}(x_{1j}, x_{2j}, \ldots, x_{nj})}{\text{sd}(x_{1j}, x_{2j}, \ldots, x_{nj})},
\end{equation}
with $x_{ij}$ being the $j^{th}$ environmental covariate for the $i^{th}$ observation.  Here, the term $A_{\bm{s}(i)}$ denotes the random effect corresponding to location $\bm{s}$ and $\bm{s}(i)$ is the notation meaning the $i$-th observation is obtained from location $\bm{s}$. By an abuse of notation, we subsequently write $A_{\bm{s}(i)}$ as $A_{\bm{s}}$. The random effects are modeled hierarchically as
\begin{align}
\nonumber A_s | R_e &\sim \mathrm{Normal}(R_e, \tau_a), \\
 R_e &\sim \mathrm{Normal}(m_e, \tau_R),
\end{align}
where $A_{\bm{s}}$ represents the random effects of site $\bm{s}$ and follows a normal distribution with mean $R_e$ and variance $\tau_a$. The random effect of eco-region $(e)$ is introduced through $R_e$, which follows a normal distribution with mean $m_e$ and variance $\tau_R$, given by
\begin{equation}
m_e = \mu + \gamma \times d_e.
\end{equation}
Here, $\mu$ is the overall mean, $\gamma$ is the coefficient for diversity $(d_e)$ introduced at the eco-region level, and $\tau_a$ and $\tau_R$ are variances across site and eco-region, respectively.

\cite{lachs2023emergent} use a spatial beta generalized linear model to model the probability of coral bleaching ($P_i$) at a given location, which is assumed to follow a beta distribution with mean $\pi_i$ and precision parameter $\theta$, where bleaching is modeled as a function of DHW for location $i$ $(\mathrm{DHW}_i)$, i.e.,
\begin{equation}
P_i \sim \text{Beta}(\pi_i, \theta)
\end{equation}
with mean and variance given by $\mathbb{E}(P_i) = \pi_i$ and $\text{Var}(P_i) = \pi_i (1 - \pi_i) / (1 + \theta)$. The relationship between the mean bleaching proportion $\pi_i$ and its covariates is modeled through a logit link as
\begin{equation}
\text{logit}(\pi_i) = \beta_0 + \beta_1 \times \mathrm{DHW}_i + U_i + \varepsilon_i,
\end{equation}
where $U_i$ are spatial random effects modeled using a zero-mean GMRF \citep{rue2005gaussian}, i.e., $U_i \sim \text{GMRF}(\bm{0}, \bm{\Omega})$, and $\varepsilon_i \sim \mathrm{Normal}(0, \sigma^2)$ are spatially independent and identically distributed nugget components across locations, $\beta_0$ denotes the intercept, $\beta_1$ denotes the slope parameter for DHW. Here, $\bm{\Omega}$ denotes the covariance matrix based on a predefined covariance structure, e.g., the Mat\'ern class, approximated using a stochastic partial differential equation-based approach proposed by \cite{lindgren2011explicit}. The term $\sigma^2$ denotes the nugget variance. Details on GMRF construction in the presence of a nugget are described in \cite{cisneros2023combined} and \cite{hazra2025efficient}.

\subsubsection{Bayesian state space models}\label{subsubsec:statespace}
An additional critical issue in coral reef research concerns the limited understanding of the extent and drivers of variability in reef community dynamics across different sites. Although it is well established that coral reef dynamics are influenced by stochastic processes, the relative contributions of among-site and within-site variability remain insufficiently characterized. \cite{allen2017among} address this gap by quantifying these sources of variability using a Bayesian vector autoregressive state-space model. While the authors propose the model for spatiotemporal datasets, we do not categorize it as a spatiotemporal model because the model explains the temporal dynamics independently across locations, and the error components are both spatially and temporally independent.

To conduct their analysis, the authors first transform the coral compositional data into isometric log-ratio coordinates, allowing for the appropriate statistical treatment of proportional data. Suppose $\bm{Z}_{i, j, t} = (z_{1, i, j, t}, z_{2, i, j, t}, z_{3, i, j, t})^\top$ is the vector of observed proportional cover of coral, algae, and other at site $i$, transect $j$, and at time $t$. The transformed data vector is represented by the vector $\bm{Y}_{i,j,t}$, which is of length two; its entries represent the natural logarithm of the algae-to-coral ratio and the natural logarithm of the ratio of other to the geometric mean of algae and coral. The authors then employ a spatial first-order vector autoregressive model to analyze the temporal evolution of reef community composition across sites. The state-space model is formulated with two hierarchical components: (a) the process layer, and (b) the observation layer. The model structure can be expressed as follows.

\paragraph{(a) Process layer:}  
The change in the true, unobserved state of reef composition at site $i$ at time $t+1$ is modeled as
\begin{equation}
\bm{X}_{i,t+1} = \bm{\alpha} + \bm{B} \bm{X}_{i,t} + \bm{A}_i + \bm{\varepsilon}_{i,t},
\end{equation}
where $\bm{X}_{i,t+1}$ denotes the vector of true transformed composition at site $i$ and time $t$, $\bm{\alpha}$ denotes the mean intercept vector representing the average change across all sites, $\bm{A}_i$ denotes the random site effect, where $\bm{A}_i \sim \mathrm{Normal}_2(\bm{0}, \bm{\Sigma}_X)$, $\bm{B}$ denotes the matrix representing the effects of the current state on the next state, $\bm{\varepsilon}_{i,t}$ denotes the random temporal variation for site $i$ at time $t$, where $\bm{\varepsilon}_{i,t} \overset{\mathrm{IID}}{\sim} \mathrm{Normal}_2(\bm{0}, \bm{\Sigma}_\varepsilon)$, and it is independent of among site variation $\bm{A}_i$. 
Here, $\bm{\Sigma}_X$ and $\bm{\Sigma}_\varepsilon$ are the covariance matrices of the among-site variation $\bm{A}_i$ and the covariance matrix of the temporal variation $\bm{\varepsilon}_{i,t}$, respectively.

\paragraph{(b) Observation layer:}  
In this layer, the model describes the relationship between the true, unobserved state $\bm{X}_{i,t}$ and the observed data $\bm{Y}_{i,j,t}$ collected from the transects.  
The observed reef state $\bm{Y}_{i,j,t}$ is modeled as
\begin{equation}\label{eq:obs_eqn}
\bm{Y}_{i,j,t} = \bm{X}_{i,t} + \bm{e}_{i,j,t},
\end{equation}
where $\bm{e}_{i,j,t} \sim t_2(0, \mathcal{V}, \nu)$ IID over $j$. Here, $t_2$ denotes the bivariate $t$-distribution, and $\mathcal{V}$ and $\nu$ denote the scale matrix and the degrees of freedom for the $t$-distribution.

\subsection{Advanced Bayesian spatiotemporal models}
\label{subsec:bayesian_spatiotemporal}

\subsubsection{Spatiotemporal models}
In ecological research, particularly in the study of coral reef health, it is often essential to integrate heterogeneous datasets derived from multiple sources to obtain a comprehensive understanding of ecosystem dynamics. However, the complexity and diversity of coral reef systems pose significant challenges for data integration. Observations are typically collected using various methodologies, at different spatial and temporal scales, and with varying levels of data quality and uncertainty, which complicates unified analyses. \cite{peterson2020monitoring} address these challenges by focusing on the integration of disparate datasets to improve monitoring of the GBR ecosystem. Data for this region are gathered by numerous organizations, including citizen science initiatives, resulting in a rich but inconsistent data landscape. The authors highlight the difficulty of merging these diverse data sources in a coherent and statistically sound manner.

To overcome these challenges, they propose a weighted spatiotemporal Bayesian hierarchical model that enables the combination of data from multiple sources while accounting for their respective uncertainties appropriately. The core innovation lies in the introduction of a weighting scheme, where datasets are assigned weights according to their data quality, collection methodology, and associated uncertainty. This framework emphasizes the use of high-quality datasets while also leveraging informative content from less reliable sources. The proposed spatiotemporal Bayesian model captures variations in coral cover across both space and time, treating the true underlying coral condition as a latent (unobserved) process. By adopting a Bayesian hierarchical structure, the model offers a principled approach to propagate uncertainty across layers of inference and combine heterogeneous data sources in a statistically rigorous manner. The model formulation is expressed as follows.

To account for differences in data quality, each coral cover measurement is assigned a weight.  
For citizen-contributed data, the weight for an image $j$ classified by a person $p$ from a source $s$ is calculated as a product of four components
\begin{equation}
w_{j,p,s} = w^{(e)}_{j,s} \, w^{(n)}_{j,p,s} \, w^{(a)}_{p} \, w^{(N)}_{s},
\end{equation}
with $w^{(e)}_{j,s} = A_j / \max(A_j)$, where $w^{(e)}_{j,s}$ denotes the weight for the physical image extent (area), $w^{(n)}_{j,p,s}$ denotes the weight for the number of classification points used in the image, $w^{(a)}_{p}$ denotes the classification accuracy of the citizen $p$ classifying the image, $w^{(N)}_{s}$ denotes a factor to account for the number of images needed to generate a single coral cover estimate, and $A_j$ denotes the physical area of the image captured by a camera.

The accuracy for citizen $p$ across a set of validation images $J_p$ is defined as
\begin{equation}
w^{(a)}_{p} = \frac{1}{|J_p|} \sum_{j \in J_p} \frac{\textrm{TP}_{j,p} + \textrm{TN}_{j,p}}{\textrm{TP}_{j,p} + \textrm{TN}_{j,p} + \textrm{FP}_{j,p} + \textrm{FN}_{j,p}},
\end{equation}
where $\textrm{TP}_{j,p}$ denotes the number of true positives identified by citizen $p$ in image $j$, $\textrm{TN}_{j,p}$ denotes the number of true negatives identified by citizen $p$ in image $j$, $\textrm{FP}_{j,p}$ denotes the number of false positives identified by citizen $p$ in image $j$, $\textrm{FN}_{j,p}$ denotes the number of false negatives identified by citizen $p$ in image $j$, $J_p$ denotes the collection of images classified by citizen $p$, and $|J_p|$ denotes the cardinality of the set $J_p$.

From citizen-science data, coral cover can be estimated using
\begin{equation}
Y_{j,p,s} = \frac{1}{w^{(n)}_{j,p,s}} \sum_{k=1}^{w^{(n)}_{j,p,s}} \mathbb{I}(Y_{j,p,s,k} = \text{``coral''})
\end{equation}
where $\mathbb{I}(\cdot)$ is the indicator function, and $Y_{j,p,s,k}$ is the classification category label for point $k$ in image $j$ from source $s$ by citizen $p$. The weighted mean of coral cover estimates from participants can then be calculated as
\begin{equation}
\bar{Y}_{j,s} = \frac{\sum_p w_{j,p,s} Y_{j,p,s}}{\sum_p w_{j,p,s}} = \frac{\sum_p w_{j,p,s} Y_{j,p,s}}{w_{j,s}},
\end{equation}
where $w_{j,s} = \sum_p w_{j,p,s}$ and $p \in P_{j,s}$, with $P_{j,s}$ which represents the collection of participants classifying image $j$ from source $s$. The estimation of coral cover $Y_{i,t,s}$ for spatial data can then be done as
\begin{equation}
Y_{i,t,s} = \frac{\sum_j w_{j,s} \bar{Y}_{j,s}}{\sum_j w_{j,s}},
\end{equation}
where $j \in J_{i,t,s}$ (the collection of images contained in cell $i$ from source $s$ and time $t$), and the total weights are given by $w'_{i,t,s} = N w_{i,t,s} / \sum_i w_{i,t,s}$, where $w_{i,t,s} = \sum_{j \in J_{i,t,s}} w_{j,s},$ with $N$ denoting the sample size. Now, the weighted spatiotemporal model is defined as
\begin{equation}
p(\bm{\theta} | \bar{Y}_{i,t,s}) \propto \exp\left[\sum{w_{i,t,s}'} L_{i,t,s'}\right],
\end{equation}
where
\begin{align}
\nonumber \bar{Y}_{i,t,s} &\sim \text{Beta}(\mu_{i,t,s}, \phi), \\
\text{logit}(\mu_{i,t,s}) &= X_{i,t} \boldsymbol{\beta} + U_i + V_t + \varepsilon_{i,t,s}, \\
\nonumber U_i &\sim \text{GMRF}(\bm{0}, \bm{\Sigma}).
\end{align}
Here, $\theta$ denotes the parameter set, $L_{i,t,s}$ is the likelihood function, $X_{i,t}$ denotes the matrix of covariates, $\boldsymbol{\beta}$ denotes the vector of regression coefficients, $V_t$ denotes the temporal effect capturing year-to-year change, modeled as a first-order random walk, $U_i$ denotes the spatial random effect modeled as a GMRF with mean $\bm{0}$ and covariance $\bm{\Sigma}$, and $\varepsilon_{i,t,s}$ is the uncorrelated error term (nugget).

The focus of spatiotemporal models is not only limited to solving data integration problems, but the model can also be used in modeling extreme events. \cite{simpson2021conditional} focus on the problem of statistical modeling of extreme events that evolve with time and space. The paper addresses the issue of limitations of models used for spatiotemporal data, which is inevitable in determining the patterns of coral extreme events. To provide a solution, they propose a new spatiotemporal model by extending the `conditional extremes' approach, which has been applied to the Red Sea SST to assess the risk of coral bleaching. The model can be viewed as follows.

\vspace{0.5cm}

\textbf{(a) Core asymptotic assumption:}
The model is built on the assumption that for a process $X(\bm{\omega})$ at a space-time location $\bm{\omega} = (\bm{s}, t)$, once it is transformed to have a standard exponential upper tail, its behavior when extreme can be described by a limiting process. Formally, conditioning on the process exceeding a high threshold $u$ at location $\bm{\omega}_0$, the rest of the process converges to a residual process $Z^0$ after normalization:
\begin{equation}
\left[
\frac{X(\bm{\omega}_i) - a_{\bm{\omega}_i - \bm{\omega}_0}\{X(\bm{\omega}_0)\}}{b_{\bm{\omega}_i - \bm{\omega}_0}\{X(\bm{\omega}_0)\}},
X(\bm{\omega}_0) - u
\ \bigg| \ X(\bm{\omega}_0) > u
\right]_{i=1,\ldots,dm}
\xrightarrow{d} 
\left[ Z^0(\bm{\omega}_i) \mid E \right]_{i=1,\ldots,dm},
\end{equation}
as $u \to \infty$. Here, $a(\cdot)$ and $b(\cdot)$ are normalizing functions, and $E$ is an independent standard exponential variable, $d$ is the number of spatial locations, and $m$ is the number of time points.  
This equation forms the mathematical foundation of the model.

\vspace{0.2cm}
\textbf{(b) Parametric model components:}

\paragraph{(i) Normalizing functions:}

They test both a separable and a more flexible non-separable form for the location function $a(\cdot)$. The non-separable model, which provided a better fit, is given by
\begin{equation}
\nonumber a_{\bm{\omega} - \bm{\omega}_0}(x) = x 
\left[
\lambda_T |t - t_0|^{2 \kappa_T} + 1
\right]^{-1}
\exp\left\{
-\frac{\lambda_S \|\bm{s} - \bm{s}_0\|^{2 \kappa_S}}
{\left(\lambda_T |t - t_0|^{2 \kappa_T} + 1\right)^{\eta \kappa_S}}
\right\}.
\end{equation}

The scale function $b(\cdot)$ is modeled as
\begin{equation}
\nonumber b_{\bm{\omega} - \bm{\omega}_0}(x) = 
\left[
1 + \{a_{\bm{\omega} - \bm{\omega}_0}(x)\}^{\beta}
\right].
\end{equation}

The parameters $(\lambda_S, \kappa_S, \lambda_T, \kappa_T, \eta, \beta)$ control the shape and decay of the extremal dependence in space and time, with $\eta$ controlling the strength of the space-time interaction.


\paragraph{(ii) The residual process:}

The residual process $Z^0(\omega)$ is modeled as a conditional Gaussian process, constructed by starting with a stationary spatiotemporal Gaussian process, $Z^0(\omega)$, with a separable power exponential covariance function
\begin{equation}
\text{Cov}(Z^0(\bm{\omega}_i), Z^0(\bm{\omega}_j)) = 
\sigma^2 
\exp \left\{
- \left( \frac{\| \bm{s}_i - \bm{s}_j \|}{\phi_S} \right)^{p_S}
\right\}
\exp \left\{
- \left( \frac{| t_i - t_j |}{\phi_T} \right)^{p_T}
\right\},
\end{equation}
where $\phi_S$ and $\phi_T$ are spatial and temporal range parameters and $p_S$ and $p_T$ are spatial and temporal smoothness parameters, respectively. The residual process $Z^0(\bm{\omega})$ is then defined as the process $Z(\bm{\omega})$ conditional on the constraint that $Z(\bm{\omega}_0) = 0$.

Apart from the conditional extremes framework of \cite{simpson2021conditional}, \cite{hazra2021estimating} introduce a semiparametric Bayesian spatiotemporal model, namely a Dirichlet Process Mixture of Low-Rank Student's $t$ processes. Incorporating Representative Concentration Pathways 4.5 and 8.5 from the Intergovernmental Panel on Climate Change, the proposed model is designed to estimate and predict high-resolution SST hotspots near the end of this century, which helps in identifying coral reefs at a higher risk of bleaching. This model, which effectively captures both the bulk and tail behavior of SST, enhances flexible modeling of spatially non-stationary and heavy-tailed dependence structures. The model identifies exceedance regions with controlled uncertainty through posterior predictive inference, providing a probabilistically sound method for detecting hotspots. The proposed model incorporates asymptotic dependence with a computationally effective Bayesian framework, which complements the conditional extremes approach, enabling more flexible inference on joint spatial extremes and long-term coral thermal stress projections.

\subsubsection{Spatial clustering models}
In biodiversity research, estimating community composition is a central objective, as patterns of species commonness and rarity within biological assemblages are often used to compare ecosystems and forecast ecological changes. \cite{piancastelli2024bayesian} present a Bayesian statistical framework for clustering coral reefs of GBR based on their community composition. The study focuses on abundance measurements of four key benthic groups collected across multiple reef sites between 2012 and 2017. The primary aim of the analysis is to identify clusters of reefs exhibiting similar compositional structures and to investigate temporal changes in these clusters, particularly across periods marked by extreme environmental disturbances. Recognizing that reefs in close geographic proximity are likely to share similar ecological characteristics, \cite{piancastelli2024bayesian} incorporate spatial dependence into their modeling framework through the integration of a Dirichlet Mixture Model coupled with a Potts distribution to model spatial clustering behavior. The proposed models can be formally expressed as follows.
\paragraph{(a) Dirichlet mixture models:}

The data, a set of compositions $\bm{p}$, is modeled as coming from a mixture of $k$ different Dirichlet distributions. Reef communities are classified into four groups, namely, hard coral, soft coral, algae, and sand. The model introduces a latent vector $\bm{z}$, where $z_j$ is the cluster assignment for reef $j$. The conditional likelihood of the data given cluster assignments $\bm{z}$ and cluster parameters $\bm{\rho}$ is
\begin{equation}
f(\bm{p} \mid \bm{z}, \bm{\rho}) =
\prod_{j=1}^{n}
\left(
\frac{\Gamma\left(\sum_{i=1}^{4} \rho_{z_j,i}\right)}
{\prod_{i=1}^{4} \Gamma(\rho_{z_j,i})}
\prod_{i=1}^{4} p_{ij}^{\rho_{z_j,i}-1}
\right).
\end{equation}
Here, $\bm{\rho}_\ell$ is the parameter vector for the Dirichlet distribution corresponding to cluster $\ell$.


\paragraph{(b) Potts distribution for spatial dependency:}

The cluster assignments $\bm{z}$ are not assumed to be independent. Instead, they follow a Potts distribution, which makes the probability of a reef’s assignment $z_j$ dependent on the assignments of its neighbors, $N(j)$. The joint probability of all cluster assignments is
\begin{equation}
\pi(\bm{z} \mid \boldsymbol{\nu}, v_0)
\propto q(\bm{z} \mid \boldsymbol{\nu}, v_0)
= \exp\left\{
\sum_{\ell=1}^{k-1} \nu_\ell n_\ell
+ v_0 \sum_{j=1}^{n} 
\left[
\sum_{i \in N(j)} \mathbb{I}(z_j = z_i)
\right]
\right\},
\end{equation}
where $\nu_\ell$ models the overall prevalence of cluster $\ell$, $v_0 > 0$ is the interaction parameter that measures the strength of spatial dependence. A higher $v_0$ means neighbors are more likely to be in the same cluster.

\subsection{Bayesian networks for causal and risk-based modeling}
\label{subsec:bayesian_causal}

From the preceding discussion, it is evident that the models previously reviewed vary considerably in their objectives, structure, and analytical focus, reflecting the inherent spatiotemporal heterogeneity of coral reef data. Despite these advancements, a key methodological challenge that remains insufficiently addressed is the integration of multiple heterogeneous data sources to effectively model the complex, nonlinear, and uncertain relationships that characterize coral reef ecosystems.

\begin{figure}[h!]
    \centering
    \includegraphics[width=0.8\textwidth]{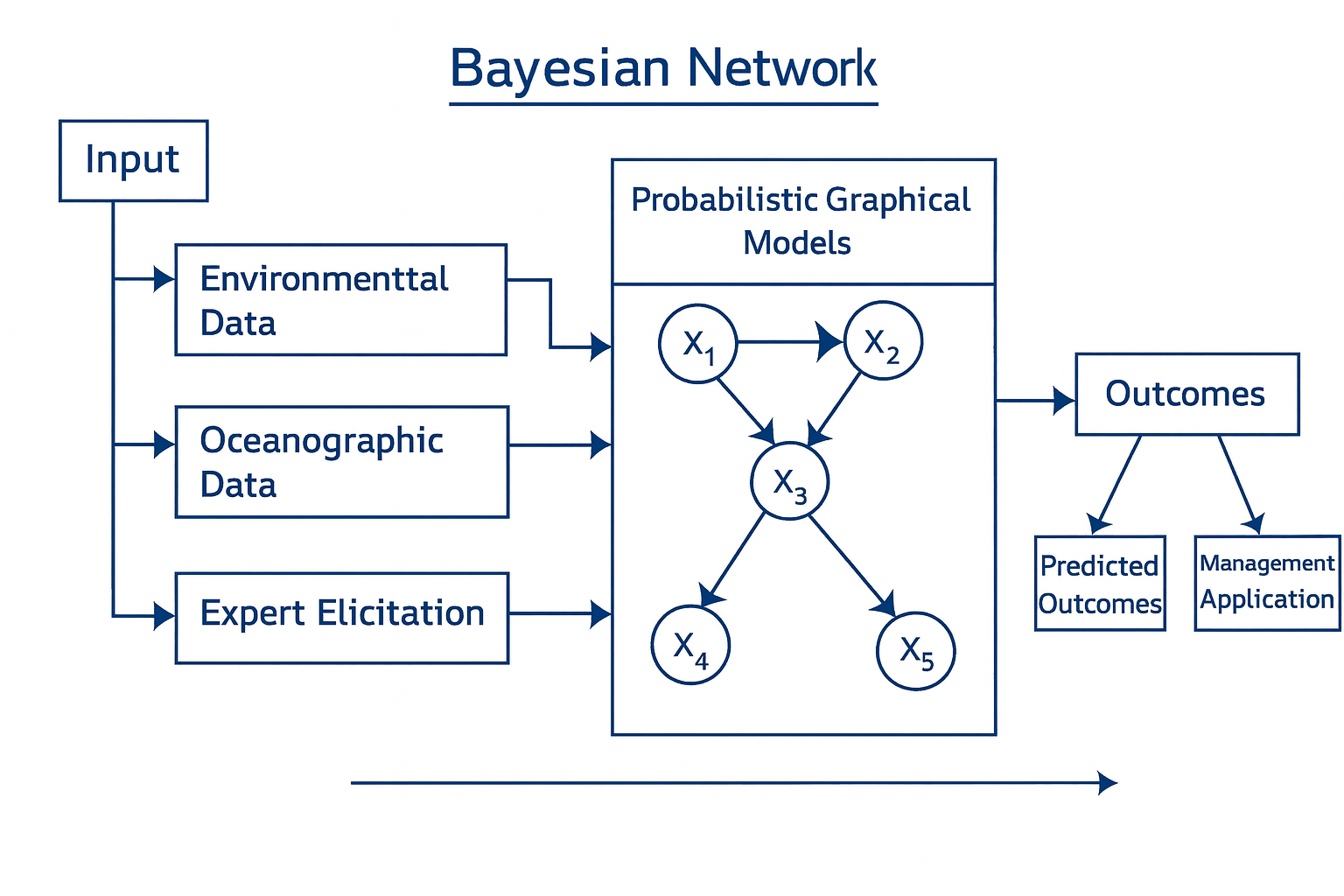}
    \caption{Bayesian Network structure showing data inputs, probabilistic graphical model, and outcomes.}
    \label{fig:Bayesian Network}
\end{figure}

In this context, Bayesian Networks have emerged as one of the most versatile and widely adopted frameworks for inferring causal and predictive relationships while accommodating multiple sources of information. Their probabilistic structure enables the fusion of diverse datasets, including in-situ measurements, remote sensing products, expert knowledge, and ecological indicators, within a coherent statistical framework. Several studies—such as \cite{bainbridge2019use}, \cite{krug2013construction}, \cite{carriger2020assessing}, \cite{ban2015assessing}, and \cite{mentzel2024evaluating}—implement Bayesian Networks for a range of purposes, including predicting coral bleaching events, quantifying the effects of multiple stressors (e.g., thermal, oceanographic, and anthropogenic), and evaluating management interventions under uncertainty. Collectively, these applications highlight the adaptability of Bayesian Networks in capturing both causal dependencies and predictive relationships across spatial and temporal scales. A generalized schematic representation of a Bayesian Network structure applicable to coral reef systems is illustrated in Figure \ref{fig:Bayesian Network}, which demonstrates how interlinked environmental, biological, and anthropogenic factors can be systematically integrated within a unified probabilistic framework.

Figure \ref{fig:Bayesian Network} illustrates the fundamental structure of a Bayesian Network. The input layer (leftmost section) represents the integration of heterogeneous information sources—such as environmental and oceanographic variables (e.g., SST, salinity), field-based ecological observations, and other relevant datasets. These inputs collectively form the basis for probabilistic inference within the network.

At the core of the framework lies the Probabilistic Graphical Model, in which nodes denote random variables and directed edges represent conditional dependencies between them. Each node is characterized by a probability distribution, capturing the uncertainty associated with that variable. Two central concepts underpin the structure of such a network: Parent nodes, denoted as Parents $X_i$ for a variable $X_i$, and CPT.

\begin{figure}[h!]
    \centering
    \includegraphics[width=0.4\textwidth]{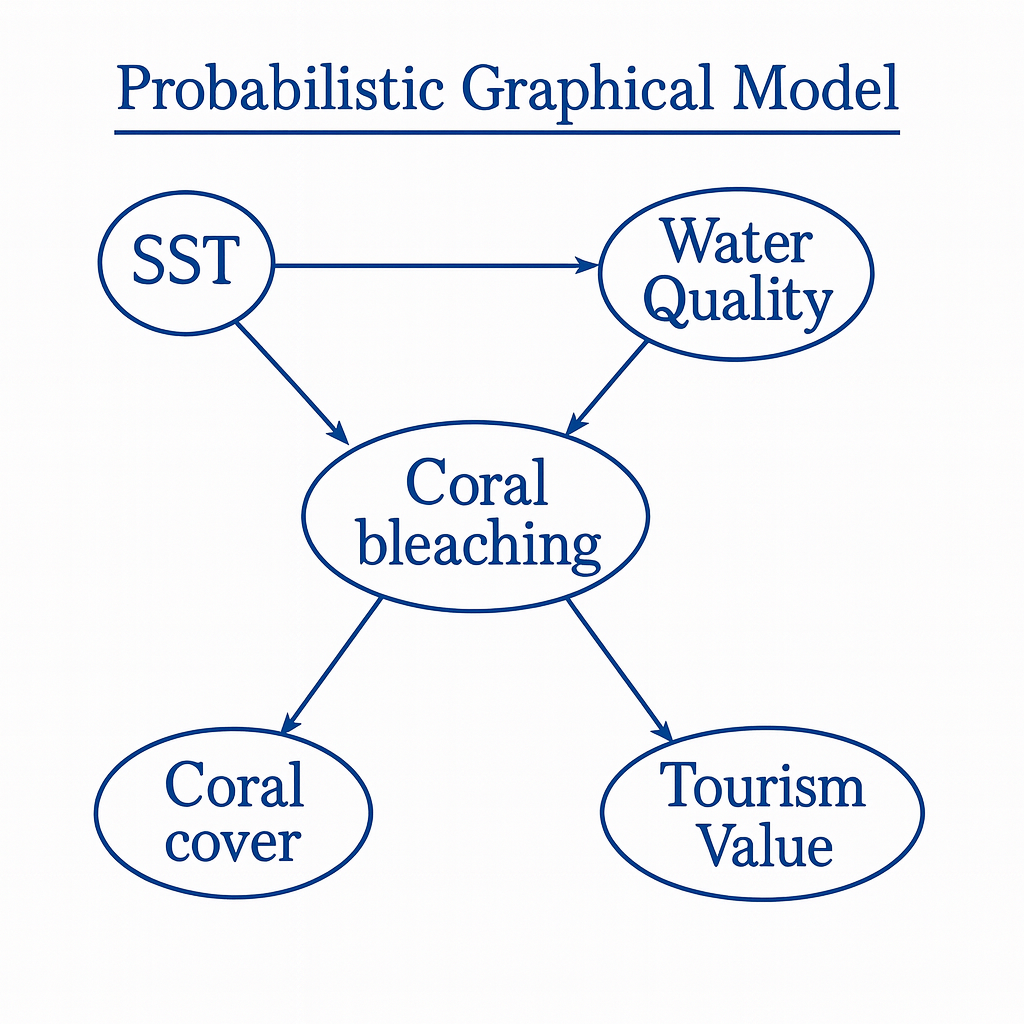}
    \caption{Probabilistic graphical model showing the influence of SST and Water Quality on coral bleaching and its impacts.}
    \label{fig:pgm}
\end{figure}

In a Bayesian Network, a parent node exerts a direct probabilistic influence on another node. For instance, in Figure \ref{fig:pgm}, SST acts as a parent to both WQ and coral bleaching, indicating that variations in SST directly affect these factors. Similarly, WQ serves as a parent node to Coral Bleaching, capturing its mediating role in the causal pathway. This structure enables the network to represent complex causal hierarchies and quantify uncertainty propagation across interconnected environmental and biological processes. 
In the illustrated network, WQ is directly influenced by SST, making SST the parent of WQ. Similarly, the variable Coral Bleaching has two parent nodes: {SST, WQ}, reflecting the combined influence of both environmental temperature and water conditions on bleaching dynamics.

The relationship between each node and its respective parents is formally represented through a CPT, which quantifies how the probability distribution of a child node depends on the states of its parent nodes. In other words, it specifies the likelihood of each possible outcome of the child variable, given the various configurations of its parent variables. This probabilistic mapping can be illustrated in the following table, which summarizes the conditional dependencies within the network.

\begin{table}[ht]
\centering
\caption{CPT for Coral Bleaching given SST and WQ. Here, LB, WB, and HB stand for low, medium, and high bleaching, respectively.}
\label{tab:cpt_bleaching}
\renewcommand{\arraystretch}{1.2} 
\setlength{\tabcolsep}{8pt} 
\begin{tabular}{l l c c c}
\hline
SST & WQ & P(LB) & P(MB) & P(HB) \\
\hline
Low      & Good & 0.80 & 0.10 & 0.10 \\
Moderate & Good & 0.55 & 0.25 & 0.20 \\
High     & Good & 0.25 & 0.35 & 0.40 \\
Low      & Poor & 0.62 & 0.27 & 0.11 \\
Moderate & Poor & 0.33 & 0.38 & 0.29 \\
High     & Poor & 0.17 & 0.33 & 0.50 \\
\hline
\end{tabular}
\end{table}

Table \ref{tab:cpt_bleaching} illustrates the conditional relationships between environmental variables and coral bleaching probabilities. For instance, when SST is high and WQ is poor, the model estimates the probability of high bleaching (HB) to be 0.5. Conversely, when SST is low and WQ is good, the probability of high bleaching decreases to 0.1. These conditional probabilities elucidate how variations in SST and WQ jointly influence coral bleaching intensity. CPT thus serves as a fundamental component of the Bayesian Network, capturing the probabilistic dependencies among variables. CPTs may be derived from empirical observations, expert elicitation, or a combination of both, allowing flexibility in integrating heterogeneous data sources.

Mathematically, the entire Bayesian Network can be represented as a unified probabilistic system composed of several interlinked conditional structures. For $n$ input variables ${X_1, X_2,\ldots,X_n}$, the joint distribution of ${X_1, X_2,\ldots,X_n}$ across the system can be written as \[
\mathrm{P}(X_1, X_2, \ldots, X_n) = \prod_{i=1}^{n} \mathrm{P}(X_i \mid \text{Parents}(X_i)).
\]

Information-theoretic measures, such as Mutual Information and Kullback-Leibler divergence, are helpful in measuring the sensitivity between nodes. Standardized total and direct effects, as well as target mean analysis, can further clarify the direction and magnitude of each relationship, while entropy reduction and cross-entropy account for nonlinear dependencies.

The last layer of this Bayesian Network is the Output layer, which draws a conclusion from the inference phase of the central layer, which can be further decomposed into two parts, viz. Predicted Outcomes, which include the probability of bleaching and the environmental risk index, as well as management applications, such as early warning forecasts. The last-edged arrow in Figure \ref{fig:Bayesian Network} represents that if new data are collected, they can be incorporated in the model to revise the estimates. This can be done via Bayes’ theorem. Bayesian Networks provide a powerful probabilistic framework for modeling coral bleaching events by integrating diverse data sources within a coherent hierarchical structure. This approach has been implemented in multiple ways across coral reef studies, reflecting its adaptability to different spatial scales, data types, and research objectives.

To identify the key drivers influencing bleaching intensity on Brazilian reefs, \cite{krug2013construction} employ causal Bayesian Networks, where variables such as temperature, light attenuation, wind, and rainfall were hierarchically linked within the network. Their probabilistic classification of weak versus strong bleaching events revealed that accumulated SST anomalies were the most dominant factors influencing bleaching outcomes \citep{lisboa2018nino, wooldridge2004learning}. For GBR, \cite{ban2015assessing} apply expert-elicited spatial Bayesian Networks to evaluate the effectiveness of local management strategies under projected climate change scenarios \citep{ban2014assessing, renken2009modelling}. The study integrated expert-derived CPTs with multiple spatial datasets, including fishing effort, sedimentation, and temperature anomalies, to simulate future conditions. The resulting probability estimates identified reef regions where enhanced local management could partially mitigate global stress effects \citep{franco2016bayesian}. Focusing on the 2015–-2016 GBR bleaching event, \cite{bainbridge2019use} develop an empirical Bayesian Network to generate daily bleaching risk indices \citep{bainbridge2014use}. By incorporating real-time sensor data on environmental variables such as light and temperature, the hierarchical model generated an overall bleaching risk node. Continuous data updates enabled the network to deliver short-term probabilistic forecasts and early-warning alerts for reef managers \citep{pearson2017bayesian}. \cite{carriger2020assessing} utilize data-driven Bayesian Networks to examine the relationships between global and local stressors and various coral condition indicators, such as live coral cover, bleaching, and mortality, across six oceanic regions \citep{carriger2024exploring}. Their approach combined an augmented Naive Bayes classifier with extensive sensitivity analyses, including Mutual Information, standardized effects, and target mean analysis, to quantify the influence of each stressor. More recently, \cite{mentzel2024evaluating} propose a multi-model Bayesian Network framework that integrates climate projections, hydrological simulations, and ecological stressor dynamics to assess the potential impacts of future climate scenarios on GBR coral health. Their findings indicate that SST and WQ are the primary determinants of bleaching intensity and that improving local WQ could help mitigate climate-induced stress. The model effectively incorporated uncertainty from both climatic and hydrological sources, demonstrating the robustness of this probabilistic approach.

Collectively, these studies illustrate how Bayesian Networks, grounded in the same fundamental probabilistic principles, can flexibly integrate heterogeneous data across spatial and temporal scales. This adaptability underscores their effectiveness in modeling the complex and multifaceted nature of coral bleaching processes.

\subsection{Advanced statistical metrics and operators}
\label{subsec:statistical_metrics}

In many cases, assessing the complex spatial and temporal dynamics of coral reef resilience is as important as finding the relationships between predictors and the coral bleaching response. It is due to a lack of robust, quantitative frameworks that would account for coral reef resilience. 

\cite{eason2024assessing} address this urgent challenge of declining coral reef resilience across the globe. Their study employs resilience-based quantitative methods to detect degradation patterns. They have used univariate indicators such as variance and skewness to assess whether system variability and asymmetry increase prior to ecological regime shifts. In multivariate indices, they use the Variance Index, which is the maximum eigenvalue of the covariance matrix of system variables, to measure the dominant variance component in the system. They also use FI, an information theory-based metric that quantifies the extent of stability in the patterns of the system. The empirical computation involves grouping system states into bins based on tolerance thresholds and then estimating the probability of observing each state, $p(s)$. Using this, FI can be calculated as
\begin{equation}
\mathrm{FI} = 4 \sum \left[ q(s) - q(s+1) \right]^2,
\end{equation}
\noindent
where $q(s) = \sqrt{p(s)}$. Here, FI being sensitive to both spatial and temporal variation helps to detect the transition from stable to unstable regimes.

\cite{van2009quantifying} propose a forecast verification method from meteorology using the Peirce Skill Score to assess and optimize the predictive quality of the DHW technique. Their study involves constructing a $2\times2$ contingency table to compare predicted and observed bleaching for each grid cell and year, as presented in Table \ref{tab:contingency}. 
\begin{table}[h!]
\centering
\caption{Contingency table for evaluating bleaching prediction performance.}
\label{tab:contingency}
\begin{tabular}{|c|c|c|}
\hline
\textbf{} & Bleaching Observed & No Bleaching Observed \\ \hline
Bleaching Predicted & $a$ (hits) & $b$ (false alarm) \\ \hline
No Bleaching Predicted & $c$ (misses) & $d$ (correct rejection) \\ \hline
\end{tabular}
\end{table}

\noindent
The computed metrics from Table \ref{tab:contingency} are as follows:
\begin{eqnarray}
\nonumber    H &=& \frac{a}{a + c} \quad \text{(Hit rate)}, \\
F &=& \frac{b}{b + d} \quad \text{(False alarm rate)}.
\end{eqnarray}
Here, the Peirce Skill Score is defined as $PSS = H-F$, where $PSS = 1$ indicates perfect prediction, $PSS = 0$ indicates constant (no-skill) prediction, and $PSS < 0$ indicates anti-correlation. This measure accounts for both false alarms and missed bleaching events, while avoiding biased forecasts.

\cite{ledrew2004spatial} address the difficulty of detecting and quantifying spatially distributed coral stress and damage from multi-spectral imagery without extensive in-situ data and correction for water depth and atmospheric effects. The authors propose an approach that is based on a change in spatial structure within reef imagery over time, as a healthy coral is expected to exhibit high spatial heterogeneity due to coral colonies, algae, and sand, whereas a degraded coral is more spatially homogeneous because of coral death and algal overgrowth, smoothing reflectance patterns. To account for the degree of spatial dependence among pixel values in the image, the study employs a spatial autocorrelation approach using Getis statistics ($G_i^*$), which is defined as
\begin{equation}
G_i^* = 
\dfrac{
    \sum_{j} w_{ij}(d) x_j - \bar{x} \, w_i^*
}{
    s \sqrt{ \dfrac{n w_i^* - (w_i^*)^2}{n - 1} }
},
\label{eq:getis}
\end{equation}
where $x_j$ is the pixel value at location $j$, $w_{ij}(d)$ is the spatial weight (1 if pixel $j$ is within distance $d$ of pixel $i$, 0 otherwise), $w_i^* = \sum_{j} w_{ij}(d)$ denotes the number of pixels in the window centered on pixel $i$, $\bar{x}$ is the mean of all pixel values in the image, $s$ is the global standard deviation of pixel values, and $n$ is the total number of pixels in the image.


The computed value of $G_i^*$ behaves like a $z$-score where high positive values indicate clustering of high reflectance, i.e., spatial homogeneity, possible bleaching dominance, etc., and high negative values indicate clusters of low reflectance, i.e., spatial heterogeneity, healthy coral patches, etc. The statistic is computed using multiple moving windows ($3\times3$, $5\times5$, $7\times7$ pixels). For each pixel, the window size that yields the maximum $G_i^*$ is called the maximum Getis distance, indicating the size of homogeneous patches within the reef. By comparing the maximum Getis distance between two image dates, one can infer whether spatial homogeneity or heterogeneity has increased over time.

\section{Machine Learning methods} 
\label{sec:machine_learning}

Previous research has demonstrated that statistical modeling of coral bleaching events presents considerable challenges due to the complexity of the underlying ecological processes. Across a vast body of literature, researchers have employed a variety of advanced statistical techniques to model such phenomena. However, coral bleaching is inherently a multifactorial and dynamic process, influenced by a combination of environmental variables, such as SST and WQ, and local anthropogenic stressors. While the statistical models discussed thus far have been applied extensively, they often struggle to capture the nonlinear, high-dimensional, and temporally dynamic nature of coral reef ecosystems.

To address these limitations, ML approaches have emerged as powerful alternatives for modeling coral bleaching events. ML techniques offer several advantages: they can effectively model complex nonlinear interactions between responses and predictors without stringent parametric assumptions; they allow for the integration of heterogeneous data sources; and they often achieve superior predictive performance compared to traditional threshold-based methods. Furthermore, recent developments in deep learning, particularly Convolutional Neural Networks, have enabled automated classification of coral health states and detection of bleaching from underwater imagery with high accuracy.

In the following two subsections, we present an overview of how machine learning techniques have been applied to predict and monitor coral bleaching events. Broadly, these methods can be categorized into two major groups: (1) supervised learning, and (2) unsupervised learning, which are discussed in detail below.

\subsection{Supervised learning algorithms}
\label{subsec:supervised}

When the outcomes, such as bleaching occurrence or intensity, are known, supervised ML techniques are employed to learn the relationship between predictor variables and observed responses. These methods are particularly valuable for predicting bleaching status or severity based on environmental and biological covariates.

To predict coral bleaching status along the southern coasts of Thailand, \cite{boonnam2022coral} apply three supervised classification algorithms: Naïve Bayes, SVMs, and Decision Trees. Their model utilized three environmental predictors—seawater pH, SST, and wind speed—to classify reefs into five categories of coral damage: perfectly luxuriant, very good, medium, slightly damaged, and completely damaged. The Naïve Bayes classifier, a probabilistic model based on Bayes' theorem, assumes conditional independence among predictors and computes the probability of class membership given the input features. The SVM model, in contrast, seeks an optimal hyperplane that maximizes the margin between classes, effectively separating different levels of coral damage. The Decision Tree classifier employs a rule-based approach, recursively splitting the dataset into subgroups based on threshold conditions (e.g., ``Is SST~$>$~$25^{\circ} \mathrm{C}$?'') until final predictions are made. Among these algorithms, the SVM achieved the highest classification accuracy (88.85\%), demonstrating superior predictive capNaïvety in this context.

Similarly, \cite{welle2017estimating} evaluate the impacts of multiple environmental stressors, such as temperature, solar radiation, depth, hurricanes, and anthropogenic activity, on coral bleaching and mortality during the 2005 Caribbean event. Although several GLMs were tested, the SVM was included as a nonlinear benchmark to assess whether machine learning approaches could outperform classical statistical models in predicting bleaching outcomes.

In another study, \cite{mcclanahan2019temperature} analyze bleaching severity using 26 predictor variables by comparing two modeling approaches: GLMs and BRTs. The BRT method is an ensemble learning technique that sequentially combines multiple simple decision trees, where each subsequent tree models the residuals (errors) of the previous one. This `boosting' process iteratively refines the predictions, yielding a highly accurate model that can detect complex, nonlinear relationships and high-order interactions. The study highlighted BRT’s ability to identify key predictors, such as hot and cold spells, without requiring explicit specification of interaction terms.

\cite{kumagai2018high} compare two predictive models, binomial GLM and RF, for forecasting bleaching events. The RF algorithm, another ensemble-based approach, constructs multiple decision trees using bootstrapped samples and random subsets of predictors. The final prediction is obtained by aggregating individual tree outputs, enhancing model robustness and reducing overfitting.

A novel approach by \cite{mayfield2023field} integrate proteomics data and machine learning to predict the bleaching susceptibility of the Caribbean coral Orbicella faveolata. The objective is to determine whether protein abundance profiles can serve as predictors of bleaching outcomes during the 2019 heat stress event. The response variable, Colony Health Designation, consists of three classes: bleaching-susceptible, bleaching-resistant, and intermediate. The authors train ANNs on proteomic features, employing a combination of activation functions (hyperbolic tangent, linear, and Gaussian) and feed-forward backpropagation algorithms. The best-performing ANN achieves an accuracy exceeding 80\%, while an SVM used for comparison achieves nearly 90\% accuracy, underscoring the predictive power of molecular-level data.

Overall, these studies demonstrate that supervised machine learning methods—including Naïve Bayes, SVMs, Decision Trees, Random Forests, BRTs, and ANNs—offer robust, flexible, and high-performing frameworks for modeling and predicting coral bleaching phenomena. By capturing nonlinearities and interactions across environmental, biological, and molecular variables, these models significantly enhance our ability to understand and predict bleaching dynamics.

\subsection{Unsupervised learning algorithms}
\label{subsec:unsupervised}

In examining the spatial and ecological heterogeneity of coral reef systems, unsupervised machine learning algorithms play a crucial role in uncovering hidden patterns, groupings, and associations within unlabeled data. These methods enable researchers to detect underlying structures in complex datasets without relying on predefined outcome variables.

\cite{boonnam2022coral} apply $k$-means clustering to categorize coral reef sites exhibiting similar environmental and bleaching characteristics. The $k$-means algorithm partitions observations into $k$ clusters by minimizing within-cluster variance, thereby identifying spatial zones characterized by distinct coral health and stress responses. Their analysis reveals six clusters, each corresponding to different environmental regimes, which collectively reflect varying levels of bleaching susceptibility. In addition to clustering, the authors employ association rule mining using the Apriori algorithm \citep{agrawal1996fast} to uncover relationships between environmental variables and bleaching outcomes. The resulting association rules indicate a strong negative relationship between SST and pH, suggesting that temperature-induced acidification significantly increases bleaching risk. These findings demonstrate the potential of unsupervised learning approaches for facilitating knowledge discovery and enhancing the mechanistic understanding of coral-environment interactions, particularly in data-rich but label-scarce ecological contexts.

\section{Future research directions}
\label{sec:future}

Several significant statistical and machine learning methods have been developed for modeling coral bleaching-related problems over the last two decades. However, some methodological and conceptual challenges remain that limit the accuracy and interpretability of these approaches. Future research on coral bleaching prediction should bridge the gap between process-based ecological understanding and the predictive power of these approaches. While traditional models, such as GLMs, GAMs, and Bayesian and spatial frameworks, have provided valuable insights into the factors of coral bleaching, they have limited use in capturing complex, nonlinear relationships and spatial dependencies, which need to be addressed. At the same time, emerging machine learning methods, such as RFs, Decision Trees, and SVMs, have shown strong predictive performance, often at the cost of interpretability. A promising future research direction lies in an interpretable machine learning framework that combines process and data-driven approaches, which would not only increase accuracy but also provide ecological explanations. Many models rely on simplifying assumptions such as spatial dependence, stationarity of relationships, or ignoring structural zeros. Addressing these issues requires considering situation-based models, which, along with prediction accuracies, would serve as flexible models that account for variability and model assumptions.

\begin{figure}[h!]
    \centering
    \includegraphics[width=0.48\textwidth]{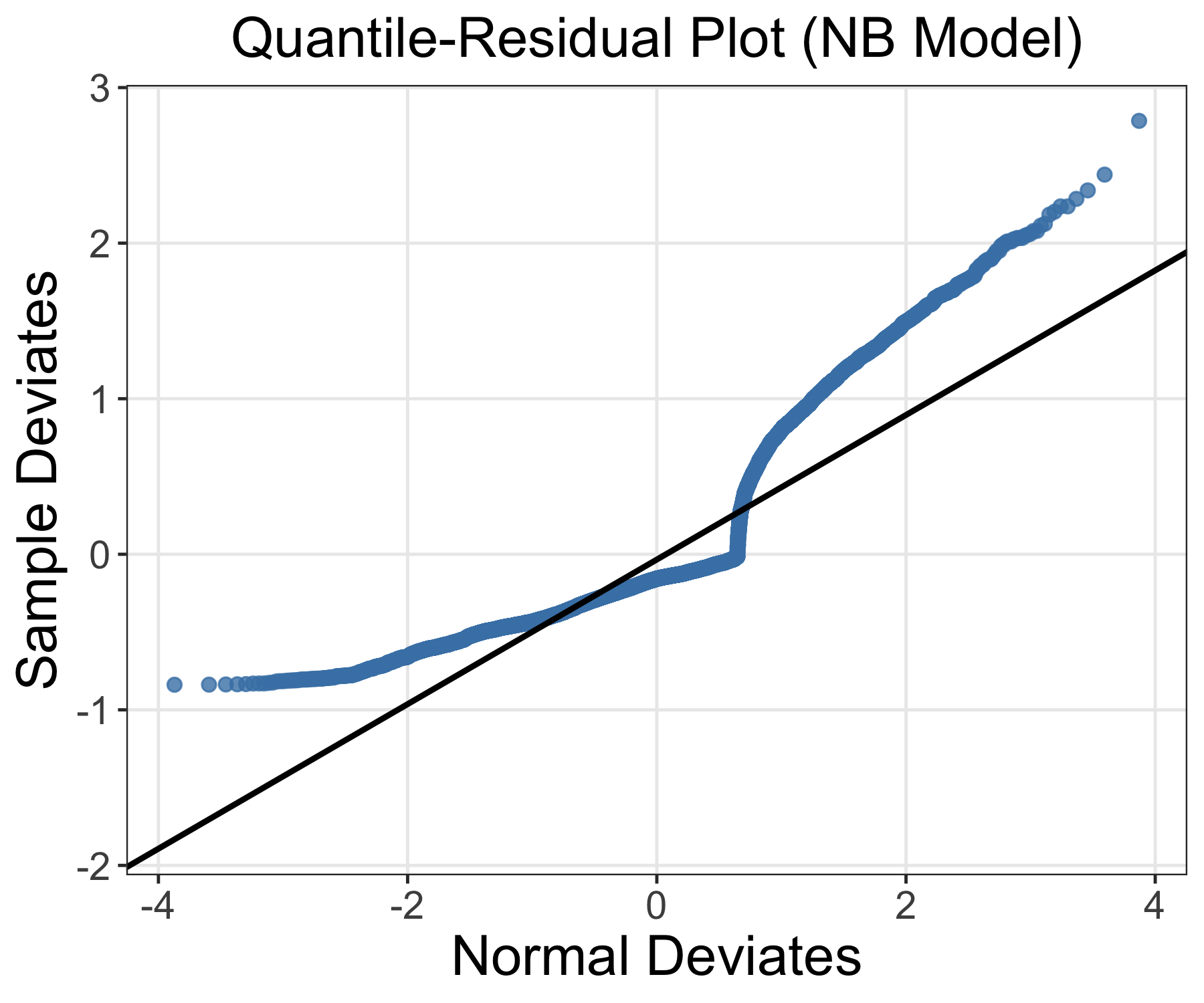}
    \includegraphics[width=0.48\textwidth]{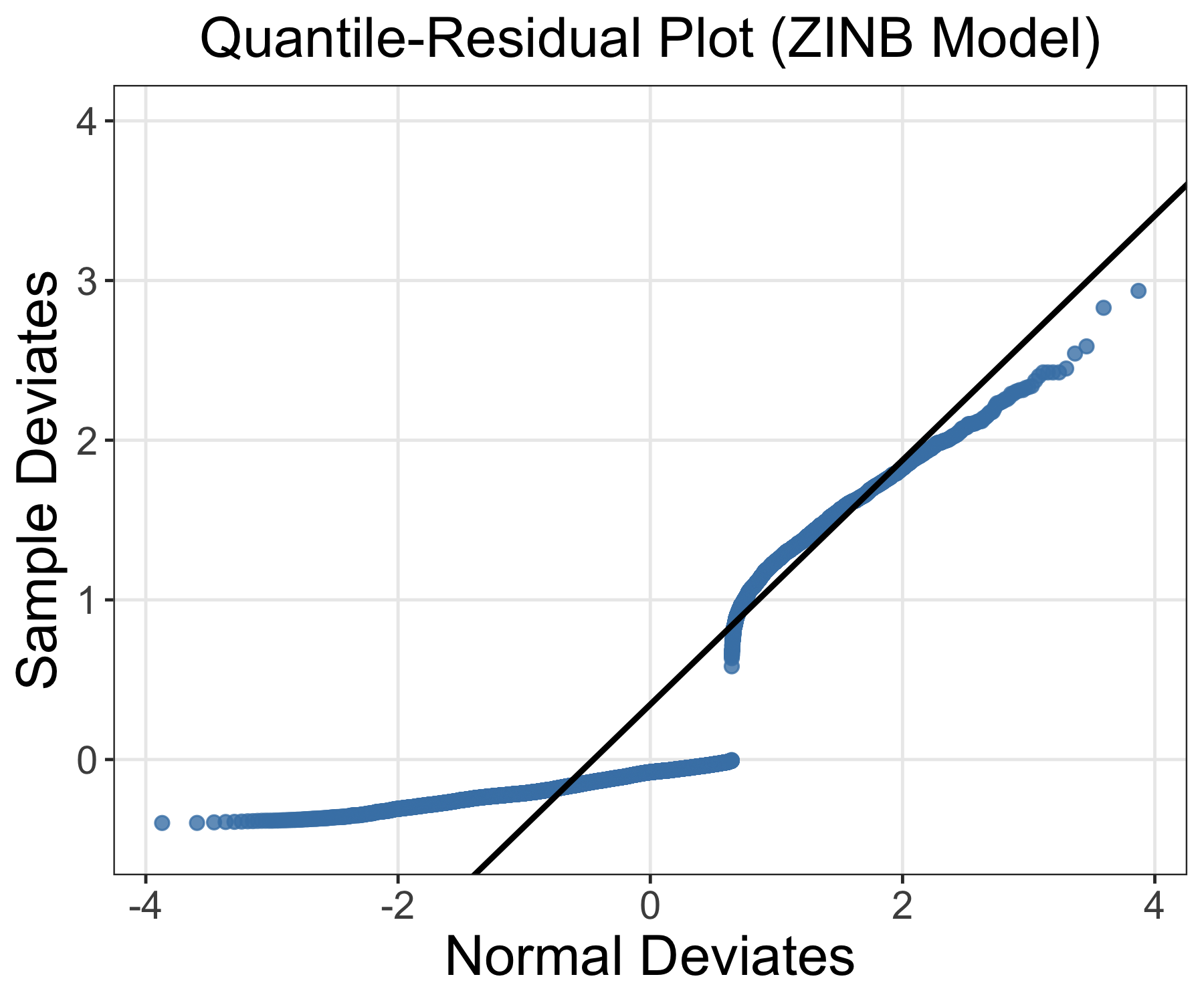}
    \caption{Left: Dunn-Smyth Residual Plot for NB Distribution. Right: Dunn-Smyth Residual Plot for ZINB Distribution.}
    \label{fig:Residual Plot}
\end{figure}

For predicting bleaching percentage, \cite{sully2019global} use a negative binomial distribution to account for zeros in their dataset with mean parameter $p_i$ and dispersion parameter $r$. From Figure \ref{fig:Residual Plot}, it is evident that the upward-bending nature of the curve \citep[Dunn-Smyth residual plot,][]{dunn1996randomized, gelman1995bayesian} and the distance from the normal line indicate that the overdispersion remains after fitting the negative binomial distribution in the Bayesian framework. The variance of the predicted mean in Table \ref{table:ppc_summary_NB} also indicates that the model is incapable of capturing the zeros in the data. 

\begin{table}[ht]
\centering
\caption{Posterior Predictive Summary for NB and ZINB distributions.}
\label{table:ppc_summary_NB}
\renewcommand{\arraystretch}{1.2} 
\setlength{\tabcolsep}{8pt} 
\begin{tabular}{l l l l}
\hline
Statistic & Observed & Fitted (NB) & Fitted (ZINB) \\
\hline
Mean & 2.45 & 3.78 & 1.72 \\
Variance & 78.90 & 1383.74 & 866.73 \\
Proportion of zeros  & 0.74 & 0.73 & 0.87 \\
\hline
\end{tabular}
\end{table}

In the coral bleaching dataset, the response variable (bleaching counts) contains a large proportion of zeros. In contrast, the model assumes that all the zeros are coming from the same stochastic process. In reality, many reefs may not be exposed to bleaching events due to their location or natural resilience, which makes them significantly different from reefs that could bleach but did not in a particular year, rendering the model incapable of accurately capturing them. Failing to account for this can bias parameter estimates and obscure key ecological processes. A suggested framework might be to use a ZINB distribution, in which a Bernoulli component models the structural observations, and an NB component models the count process for susceptible sites. Replacing the model \eqref{eq:NB} (without changing the hierarchical layer of the random effects) with a ZINB distribution can be described as follows.

Let a zero-inflation probability be $\phi_i \in [0,1]$ and $Z_i \sim \text{Bernoulli}(\phi_i),$
where $Z_i$ is an indicator for a structural zero. Then, we model the data as the observed count $Y_i =0$ if $Z_i=1$ and $Y_i=Y_i^*$ if $Z_i=0$, where $Y_i^* \sim \text{NB}(p_i, r)$ with mean parameter $p_i$ and dispersion parameter $r$. Here,
\[
\mathbb{E}(Y_i) = (1 - \phi_i)p_i,~~\operatorname{Var}(Y_i) = (1 - \phi_i)\left( p_i + \frac{p_i^2}{r} \right) + \phi_i(1 - \phi_i)p_i^2.
\]

We conduct an analysis similar to \cite{sully2019global} but using the ZINB distribution. We observe that the overdispersion issue is significantly resolved after fitting the ZINB model (Figure \ref{fig:Residual Plot}), and the variance of the predicted mean also decreases to a significant extent compared to the negative binomial model, enhancing the model flexibility (Table \ref{table:ppc_summary_NB}). Moreover, the higher predicted zero proportion under ZINB supports the fact that many reefs might remain unbleached, providing a more realistic and interpretable framework for understanding the causes of coral bleaching. While this study demonstrates the improvement achieved through the ZINB model for count data with structural zeros, future research can explore other statistical frameworks to address distributional challenges, nonlinearity, and spatial complexity in coral bleaching data. 

\cite{lachs2023emergent} model bleaching intensity as a continuous proportion using a spatial beta regression model with GMRF priors. Since bleaching proportion data are bounded between zero and one and often contain a large number of zeros, a Zero-Inflated Beta model may be a suitable alternative, which would account for no bleaching events (structural zeros) at some sites. Such a framework could better capture variability in reef-level bleaching prevalence.

\cite{peterson2020monitoring} propose a weighted spatiotemporal Bayesian hierarchical model for spatially structured bleaching count data bounded by total coral observations. Extending from their framework, a Bayesian beta-binomial hierarchical model could also be considered here, enabling the joint modeling of bounded bleaching counts and spatial-temporal correlation across reefs. Such an approach could be beneficial in enhancing uncertainty quantification and yielding more stable predictions in data-sparse regions.

Furthermore, \cite{donner2005global} use a linear regression model via a statistical downscaling approach to assess coral bleaching, which could potentially be replaced by a Bayesian hierarchical downscaling model to allow location-specific regression parameters and quantify spatial uncertainty \citep{peterson2020monitoring}. Alternatively, as a flexible alternative to the linear downscaling in \cite{donner2005global}, GAMs can be used. Together, these approaches offer promising extensions to the linear framework of \cite{donner2005global}, thereby improving predictive performance and ecological interpretability.

Under the setup described in \cite{allen2017among}, each point within a transect produces a single categorical observation classified as `coral', `algae', or `other'. Let the true underlying probabilities at site $i$, transect $j$, and time $t$ be $p_{1,i,j,t} = \textrm{P}(\text{coral})$, $p_{2,i,j,t} = \textrm{P}(\text{algae})$, and $p_{3,i,j,t} = \textrm{P}(\text{other})$, with the constraint $p_{1,i,j,t} + p_{2,i,j,t} + p_{3,i,j,t} = 1$. Each observation $\bm{Z}_{i,j,t}$ in Section \ref{subsubsec:statespace} can then be modeled as
\[
\bm{Z}_{i,j,t} \mid \bm{p}_{i,j,t}
\sim \text{Multinomial}\big(1;\, p_{1,i,j,t},\, p_{2,i,j,t},\, p_{3,i,j,t}\big),
\]
where $Z_{i,j,t} \in \{(1,0,0),\, (0,1,0),\, (0,0,1)\}$ indicates whether the sampled point was classified as coral, algae, or other, and $\bm{p}_{i,j,t}=(p_{1,i,j,t},\, p_{2,i,j,t},\, p_{3,i,j,t})^\top$. When $N_{i,j,t}$ points are sampled within a transect, the aggregated counts satisfy $\bm{Z}_{i,j,t} \mid \bm{p}_{i,j,t}
\sim \text{Multinomial}\big(N_{i,j,t};\, p_{1,i,j,t},\, p_{2,i,j,t},\, p_{3,i,j,t}\big)$. To avoid the complications due to a multinomial distribution assumption at the observation layer in \eqref{eq:obs_eqn}, \cite{allen2017among} consider isometric log-ratios and model them in a bivariate Student $t$-distribution framework. Adopting this multinomial observation model would ensure that the data adhere to the compositional constraints inherent in benthic cover measurements and would provide a more principled observation layer in a similar framework. Additionally, an appropriate Markov chain Monte Carlo algorithm needs to be designed for the posterior inference.

The integration of advanced statistical frameworks such as zero-inflated or spatially explicit models offers a promising direction for coral bleaching research. Integrating them can enhance the predictive ability of the model, uncovering the complex, nonlinear interactions between environmental stressors that traditional methods might overlook. Coral reef systems continue to experience gradual environmental change, which necessitates the development of more robust and interpretable modeling frameworks for accurate bleaching forecasts and effective conservation planning.

\section{Conclusions}
\label{sec:conclusion}

Coral bleaching represents one of the most pressing challenges to marine ecosystems, driven primarily by the intensifying effects of climate change. Over the past two decades, researchers have developed a diverse range of statistical and ML approaches to model, detect, and predict bleaching events across multiple spatial and temporal scales. Traditional frequentist frameworks, including regression-based techniques such as GLMs and GAMs, have played a central role in understanding the relationships between environmental and anthropogenic stressors and bleaching intensity \citep{welle2017estimating, mcclanahan2019temperature}. In contrast, Bayesian hierarchical and spatial models treat uncertainty in a more rigorous way. They provide calibrated probabilistic inference and more precise ecological interpretation by considering both aleatoric and epistemic uncertainties in the model \citep{allen2017among, peterson2020monitoring}.

To better capture the complex, hierarchical, and nonlinear dependencies inherent in coral reef systems, researchers have increasingly turned to machine learning methods such as RF, BRT, SVM, and ANN \citep{kumagai2018high, boonnam2022coral, mayfield2023field}. These models have demonstrated superior predictive performance and scalability compared to conventional statistical models, mainly due to their ability to handle high-dimensional data and uncover intricate, nonlinear patterns without strong parametric assumptions. The integration of heterogeneous data sources, including remote sensing products, in-situ observations, and omics-level data, further enhances the predictive power and transferability of these models across ecosystems and regions \citep{bainbridge2019use, carriger2020assessing}.

Despite these advances, several challenges persist. The limited availability of long-term, high-resolution, spatially standardized datasets continues to constrain model generalization and validation \citep{sully2019global, peterson2020monitoring}. The black-box nature of the ML algorithms limits interpretability and hinders mechanistic ecological insights essential for conservation \citep{peterson2020monitoring}. Addressing these challenges will require the development of data-specific and transferable models that can operate across scales while maintaining transparency and interpretability. The emergence of hybrid statistical–machine learning frameworks, which integrate mechanistic ecological understanding with data-driven predictive strength, offers a promising direction for improved inference and uncertainty quantification \citep{bainbridge2019use, peterson2020monitoring}. As coral reef systems exhibit nonlinear and regime-shifting responses to environmental stressors, equal importance should be given to data-driven, scalable, and transferable models that account for ecological, spatial, and temporal scale shifts \citep{donner2005global, lachs2023emergent}.

Future advancements may focus on combining biophysical forecasts with socioeconomic and policy factors to create scenario-based management tools, which would help connect reef vulnerability assessments with adaptive governance and conservation priorities \citep{johnson2023global}. Further, crucial steps include implementing open-access data platforms, standardized monitoring protocols, and interpretable monitoring modeling frameworks to enhance cross-institutional data sharing, model reproducibility, and transparency. The future of mathematical modeling of coral bleaching hinges on the integrated application of statistical inference, machine learning, and ecological theory. The transition towards the multi-scalar, uncertainty-aware predictive systems would enhance the understanding of coral bleaching modeling, in terms of precision and ecological implications. This synthesis enables the development of data-driven, uncertainty-aware management strategies that are essential for safeguarding coral reef ecosystems and sustaining marine biodiversity in the face of accelerating environmental change.

\bibliographystyle{CUP}
\bibliography{References}

\end{document}